\documentclass[aps,prb,twocolumn,showpacs,amsmath,amssymb,amsthm,superscriptaddress,groupedaddress]{revtex4}
\usepackage{graphics}
\usepackage{epsfig}
\usepackage[colorlinks=true,citecolor=blue,linkcolor=red,linktocpage=true,pagebackref=false]{hyperref}
\usepackage{soul} 
\usepackage[usenames, dvipsnames]{color}
\usepackage{braket}

\begin{document}
\title{
Single-electron $G^{(2)}$ function at nonzero temperatures 
}
\author{Michael Moskalets}
\email{michael.moskalets@gmail.com}
\affiliation{Department of Metal and Semiconductor Physics, NTU ``Kharkiv Polytechnic Institute", 61002 Kharkiv, Ukraine}

\date\today
\begin{abstract}
The single-particle state is not expected to demonstrate second-order coherence. 
This proposition, correct in the case of a pure quantum state, is not verified in the case of a mixed state. 
Here I analyze the consequences of this fact for the second-order correlation function, $G ^{(2)}$, of electrons injected on top of the Fermi sea with nonzero temperature. 
At zero temperature, the function $G ^{(2)}$ unambiguously  demonstrates whether the injected state is a single- or a multi-particle state:  
$G^{(2)}_{}$ vanishes in the former case, while it does not vanish in the latter case. 
However, at nonzero temperatures, when the quantum state of injected electrons is a mixed state, the purely single-particle contribution makes the  function $G ^{(2)}_{}$ to be non vanishing even in the case of a single-electron injection. 
The single-particle contribution puts the lower limit to the second-order correlation function of electrons injected into conductors at nonzero temperatures. 
The existence of a single-particle contribution to $G ^{(2)}_{}$ can be verified experimentally by measuring the cross-correlation electrical noise.  
\end{abstract}
\pacs{73.23.-b, 73.22.Dj, 72.10.-d, 72.70.+m}
\maketitle

\section{Introduction}

{\it Quantum coherent electronics} \cite{Splettstoesser:2009im,Haack:2013ch}, also known as electron quantum optics \cite{Bocquillon:2013fp}, and single electron electronics \cite{Bauerle:2018ct}, is an actively developing platform for quantum information processing \cite{Bennett:2000kl}, which is aimed at creating, manipulating, and detecting individual electrons as carriers of information.

Recently, quite a lot of single-electron sources on-demand have been experimentally realized.\cite{Blumenthal:2007ho,Feve:2007jx,Kaestner:2008gv,{Fujiwara:2008gt},{Roche:2013jw},Dubois:2013ul,{Rossi:2014kp},{Tettamanzi:2014gx},{dHollosy:2015ez},{vanZanten:2016fl},{Gabelli:2016uz},{Rossi:2018wj}}  
One of the crucial tests, this source has to pass through, is the verification of a single-particle emission regime.   

In quantum optics, the single-photon emission regime is verifies  via the measurement of the second-order correlation function, $g ^{(2)}$, which characterizes the  probability of joint detection of two photons.\cite{Walls2008}
Such a verification is universal and it does not rely on any specific properties of the source. 
If the stream generated by a periodically working source consists of non overlapping  single photons, then the function $g ^{(2)}\left( \tau \right)$ vanishes at zero time delay between the two detections, $ \tau=0$. 
In contrast, if there are multi-photon wave packets in the stream, the two photons can be detected simultaneously and the function $g ^{(2)}_{}$ is finite at $ \tau=0$.

The measurement of the joint detection probability in the optical frequencies range is possible due to availability of efficient single photon detectors. 
In the microwave frequencies range, no efficient detectors are available. 
Nevertheless, the single-particle emission regime for a source of microwave photons \cite{Houck:2007dy} can be demonstrated via the linear amplification of the magnitude of an electromagnetic field \cite{Bozyigit:2010bw}.

There are no efficient on-fly detectors available for single electrons so far, and there is no way to measure the magnitude of a fermionic field. 
This is why for the verification of a single-electron emission regime the various  nonuniversal methods were used. 
The nonuniversality in this context means that the given method can be good for one system, but not for another. 
In particular, a strong decrease in an electrical noise was used as an indicator of the single-particle emission regime for a dynamical quantum dot \cite{{Maire:2008hx}} and for a quantum capacitor \cite{{Mahe:2010cp},{Albert:2010co},{Parmentier:2012ed}}, while this method does not work in the case of the source of levitons\cite{Dubois:2013ul}.   
Another method for validation of the single-electron injection regime, which is relied on the partition noise\cite{{Reznikov:1998kn},Blanter:2000wi} of an electron beam splitter, was demonstrated in Refs.~\onlinecite{Bocquillon:2012if} and \onlinecite{Dubois:2013ul}.  

Nevertheless, in some systems it is possible to measure directly the second-order correlation function for injected electrons, $G ^{(2)}_{}$, which vanishes identically in the case of a single-particle injection.  
Generally, in the case of electrons injected into an electron waveguide, the second order correlation function contains several contributions: (i) One is due to electrons belonging to the Fermi sea of the waveguide, (ii) one more is due to the injected electrons, that is $G ^{(2)}_{}$, (iii) and, finally, the last contribution is due to the joint contribution of the injected electrons and Fermi-sea electrons. \cite{Moskalets:2014ea,Thibierge:2015up}  
As it is pointed out in Ref.~\onlinecite{Thibierge:2015up}, when electrons are injected into one of the two incoming channels of an electron beam splitter, the cross-correlation noise of currents after the beam splitter is directly related to the function $G ^{(2)}_{}$. 
The Fermi sea electrons do not contribute to the cross-correlation noise either directly or in conjunction with injected electrons. 
For this to be true, the two conditions must be met. 
First, the Fermi seas in both incoming channels have the same temperature and the same chemical potential. 
Second, the incoming and outgoing channels are spatially separated, which  can be achieved using chiral or helical edge states \cite{Buttiker:2009bg} as electron waveguides. 

Here I focus on the effect of temperature on the second-order correlation function, $G ^{(2)}_{}$, of electrons injected on top of the Fermi sea in conductors. 
The fact, that at nonzero temperatures the quantum state of injected electrons is a mixed state,\cite{Moskalets:2015ub,Moskalets:2017vk} leads to existence of a purely single-particle contribution to the correlation function $G ^{(2)}_{}$. 
This contribution puts the lower limit to the second-order correlation function, and it must be taken into account when the function  $G ^{(2)}_{}$ is used to distinguish single-electron and multi-electron quantum states. 

The paper is organized as follows: 
In Sec.~\ref{sec2} I discuss in detail how the correlation function $G ^{(2)}_{}$ changes when a pure quantum state becomes a mixed quantum state. 
In Sec.~\ref{sec3} the relation between the function $G ^{(2)}_{}$ and the current correlation function is given in frequency domain. 
The temperature dependence of the functions $G ^{(2)}$ for single- and two-electron excitations are contrasted in Sec.~\ref{sec4}. 
The conclusion is given Sec.~\ref{sec5}. 
Some details of calculations are preseted in the Appendixes \ref{ap01} and \ref{ap02}.

\section{Correlation function of electrons injected at nonzero temperatures}
\label{sec2}

A convenient quantity for characterizing the excitations injected by the  electron source into the electron waveguide is the excess first-order correlation function, $G ^{(1)}_{}$.\cite{Grenier:2011js,Grenier:2011dv,Moskalets:2013dl} 
To get rid of the contribution of the underlying Fermi sea and keep track of the contribution of injected electrons only, this function is defined as the difference of the two terms, evaluated with the source being switched on and off, respectively, $G ^{(1)}\left( 1;2 \right) = \langle \hat\Psi^{\dag}_{ }(1)  \hat\Psi_{ }(2) \rangle_{on} - \langle \hat\Psi^{\dag}_{ }(1)  \hat\Psi_{ }(2) \rangle_{off}$. 
Here $\hat\Psi^{}_{ }(j)$ is	an electron field operator in second quantization evaluated at time $t_{j}$ and point $x_{j}$, $j = 1, 2$, after the source. 
The quantum statistical average, $\langle \dots \rangle$, is performed over the state of electrons in the waveguide before the source. 
In this work I suppose that the waveguide is one-dimensional, and before the source an electron system is in equilibrium state, which is characterized by the Fermi distribution function with a temperature $ \theta$ and a chemical potential $ \mu$. 
Since I am interested in time dependence, rather than spatial dependence, below I keep argument $t_{j}$ only.

Note, the function $G ^{(1)}_{}$ for a stream of identically prepared separated electrons in a ballistic conductor was measured in Ref.~\onlinecite{Jullien:2014ii} using the tomography protocol suggested in Ref.~\onlinecite{Grenier:2011dv}. 

For noninteracting electrons, the first-order correlation function  determines the higher-order correlation functions through the corresponding Slater determinants.  
For example, the second-order correlation function,  $G ^{(2)}\left( t_{1}, t_{2}; t_{3}, t_{4} \right)$, is determined as follows,

\begin{eqnarray}
G ^{(2)}\left( t_{1},t_{2};t_{3},t_{4} \right) = 
\det 
\begin{pmatrix} 
G^{(1)}\left( t_{1};t_{4} \right) & 
G^{(1)}\left( t_{1};t_{3} \right)  \\ 
G^{(1)}\left( t_{2};t_{4} \right) & 
G^{(1)}\left( t_{2};t_{3} \right) \end{pmatrix} .
\label{gn01} 
\end{eqnarray}
\ \\ \indent
First let us consider a pure quantum state. 
For a single-particle state ($N=1$) with wave function $ \Psi_{1}\left( t \right)$, the first-order correlation function is factorized into the product of two terms that depend on one time each,  $G_{N=1} ^{(1)}\left(  t_{1}; t_{2}\right) = \Psi_{1}^{*}\left( t_{1} \right) \Psi_{1}\left( t_{2} \right)$. 
Apparently, that in this case  the second order-correlation function vanishes identically,  $G_{N=1} ^{(2)}=0$. 
However, already for a two-particle state ($N=2$), when  $G_{N=2} ^{(1)}\left( t_{1}; t_{2} \right) = \sum_{j=1}^{2}  \Psi_{j}^{*}\left( t_{1} \right) \Psi_{j}\left( t_{2} \right)$, the second-order correlation function is not zero. 
It is represented as follows, $G_{N=2} ^{(2)}\left( t_{1},t_{2};t_{3},t_{4} \right) = \Psi_{1,2}^{(2) *}\left( t_{1},t_{2} \right) \Psi_{1,2}^{(2) }\left( t_{4},t_{3} \right)$, where the two-particle wave function,  

\begin{eqnarray}
\Psi_{1,2} ^{(2)}\left( t_{1},t_{2} \right) = 
\det 
\begin{pmatrix} \Psi^{}_{1}\left( t_{1}\right) & \Psi^{}_{2}\left( t_{1} \right)  \\ \Psi^{}_{1 }\left( t_{2} \right) & \Psi^{}_{2}\left( t_{2} \right) \end{pmatrix} ,
\label{gn02}
\end{eqnarray}
\ \\ \noindent
is the Slater determinant composed of wave functions of both particles, $ \Psi_{1}$ and $ \Psi_{2}$. 

In contrast, in the case of a mixed state, the above relations become essentially modified. 
The quantum state is now characterize by the density matrix rather than the wave function. 

As an example, I consider a mixed state characteristic for electrons injected on top of the Fermi sea at finite temperatures.\cite{Moskalets:2017vk}  
The components of such a mixed state are parametrized by a continuous variable, the energy $ \epsilon$, with the probability density  $p _{ \epsilon} = -  \partial f\left(  \epsilon \right)/  \partial \epsilon$, where $f\left(  \epsilon \right) = \left(  1 + e^{\frac{  \mu + \epsilon }{ k_{B} \theta }} \right)$ is the Fermi distribution function, $k_{B}$ is the Boltzmann constant.
So, for the mixed two-particle state, the first-order correlation function reads,\cite{Moskalets:2017vk} 

\begin{eqnarray}
G_{N=2} ^{(1)}\left( t_{1}; t_{2} \right) = 
\int _{}^{ } d \epsilon p_{ \epsilon} 
\sum_{j=1}^{2}  \Psi_{j \epsilon}^{*}\left( t_{1} \right) \Psi_{j \epsilon}\left( t_{2} \right) .
\label{gn03}
\end{eqnarray}
\ \\ \noindent 
Accordingly to Eq.~(\ref{gn01}),  the second-order correlation function becomes, 

\begin{eqnarray}
G_{N=2} ^{(2)}\left( t_{1},t_{2};t_{3},t_{4} \right) = 
\int _{}^{ } d \epsilon p_{ \epsilon} 
\int _{}^{ } d \epsilon ^{\prime} p_{ \epsilon ^{\prime}} 
\Bigg\{  
\nonumber \\
\Psi_{1 \epsilon, 2 \epsilon ^{\prime}}^{(2) *}\left( t_{1},t_{2} \right) 
\Psi_{1 \epsilon, 2 \epsilon ^{\prime}}^{(2) }\left( t_{4},t_{3} \right) 
\label{gn04} \\
+ \sum_{j=1}^{2} 
\Psi_{ j\epsilon, j\epsilon ^{\prime}}^{*}\left( t_{1},t_{2} \right) 
\Psi_{ j\epsilon, j\epsilon ^{\prime}}^{ }\left( t_{4},t_{3} \right)
\Bigg\} .
\nonumber 
\end{eqnarray}
\ \\ \noindent
Here the two-particle wave function $\Psi_{1 \epsilon, 2 \epsilon ^{\prime}}^{(2) }$ is determined by Eq.~(\ref{gn02}) with $ \Psi_{1}$ being replaced by $ \Psi_{1 \epsilon}$ and $ \Psi_{2}$ being replaced by $ \Psi_{2 \epsilon ^{\prime}}$.  
In addition, we have a new function $\Psi_{ j\epsilon, j\epsilon ^{\prime}}$ dependent of two times, which is determined by the Slater determinant composed of different components of the same single-particle mixed state, 

\begin{eqnarray}
\Psi ^{}_{ j\epsilon, j\epsilon ^{\prime}}\left( t_{1},t_{2} \right) = 
\frac{ 1 }{ \sqrt{2} }
\det
\begin{pmatrix} \Psi^{}_{j\epsilon}\left( t_{1}\right) & \Psi^{}_{j \epsilon ^{\prime}}\left( t_{1} \right)  \\ \Psi^{}_{j \epsilon }\left( t_{2} \right) & \Psi^{}_{j \epsilon ^{\prime}}\left( t_{2} \right) \end{pmatrix} .
\label{gn05}
\end{eqnarray}
\ \\ \noindent
I name it {\it the two-time wave function}.
Note, that at coincident times, $t_{1}=t_{2}$, this function is zero, $\Psi ^{}_{ j\epsilon, j\epsilon ^{\prime}}\left( t_{},t_{} \right)=0$, which is a manifestation of the fermionic nature of an electron.

The contribution to  $G ^{(2)}$ due to the two-time wave function is present even in the case of a single-particle, but {\it mixed} state $\left(N=1 \right)$,

\begin{eqnarray}
G_{N=1} ^{(2)}\left( t_{1},t_{2};t_{3},t_{4} \right) &=& 
\int _{}^{ } d \epsilon p_{ \epsilon} 
\int _{}^{ } d \epsilon ^{\prime} p_{ \epsilon ^{\prime}} 
\label{gn06} \\
&&
\Psi_{ 1\epsilon, 1\epsilon ^{\prime}}^{*}\left( t_{1},t_{2} \right) 
\Psi_{ 1\epsilon, 1\epsilon ^{\prime}}^{ }\left( t_{4},t_{3} \right) .
\nonumber 
\end{eqnarray}
\ \\ \noindent
Note, at zero temperature the probability density becomes the delta function of energy, $p_{ \epsilon} = \delta\left( \epsilon - \mu \right)$, and the only component with Fermi energy, $ \epsilon = \mu$, survives.  
Since $\Psi ^{}_{ j \mu, j \mu}\left( t_{1},t_{2} \right) =0$, the second-order correlation function vanishes, $G_{N=1} ^{(2)}=0$, at zero temperature, as expected for a (pure) single-particle state.   

In contrast, at nonzero temperatures, when $p_{ \epsilon} \ne \delta\left( \epsilon - \mu \right)$, a single-particle state demonstrates some degree of second-order coherence, which is quantified by $G ^{(2)}_{N=1} \ne 0$. 
This is somewhat counter-intuitive, since the quantities like $G ^{(2)}$ are considered essentially multi-particle in nature. 

To resolve this seeming paradox, let us recall the physical meaning of  $G ^{(2)}$, specifically with pairwise equal arguments, $t_{1}=t_{4}$ and $t_{2}=t_{3}$. 
For a pure state, it is represented by the square of a two-particle wave function, $G ^{(2)}\left( t_{1}, t_{2}; t_{2}, t_{1} \right) = \left | \Psi^{(2)}\left( t_{1}, t_{2} \right) \right |^{2}$. 
The conventional meaning of the wave function square is the detection probability, the probability of a strong, projective measurement.  
In our case, it is the joint probability of two detections, at time $t_{1}$ and at time $t_{2}$. 

In the case of a state with two particles, say, with wave functions $ \Psi_{1}(t)$ and $ \Psi_{2}(t)$, both detections are possible. 
Let us suppose that in the first measurement we detect a particle with wave function $ \Psi_{1}$ at time $t = t_{1}$. 
The projective measurement means, that after the detection the wave function is changed, it is reduced to the delta function, $ \Psi_{1}(t) \sim \delta\left( t - t_{1} \right)$. 
Therefore, the original wave function cannot be measured at any other times. 
However, there is a second particle with wave function $ \Psi_{2}(t)$, which can be detected in the second measurement, say, at time $t= t_{2} \ne t_{1}$. 
Hence, we are able to perform two measurements, and the probability for such a joint measurement is given by  $G_{N=2} ^{(2)}\left( t_{1}, t_{2}; t_{2}, t_{1} \right) \ne 0$. 

The case with a single-particle state is different. 
As I mentioned above, we can perform a projective measurement on a single particle state only once, say, at $t= t_{1}$, and cannot measure it again $t= t_{2} \ne t_{1}$. 
This fact is manifested as  $G_{N=1} ^{(2)}=0$.

However, this logic fails in the case of a mixed state. 
The reason for this is that  a particle in a mixed state can be in several quantum states, components of a mixed state, appearing with some  probabilities. 
In Eqs.~(\ref{gn06}) and (\ref{gn05}) these states are $ \Psi_{1 \epsilon}$ for various $ \epsilon$.  
When we detect a particle at time $t = t_{1}$, we detect it in some particular component state, say, in the state with $ \epsilon = \epsilon_{0}$. 
As a result this component is reduced to the delta function, $ \Psi_{ 1\epsilon_{0}}(t) \sim \delta\left( t - t_{1} \right)$, and the original wavefunction cannot be measured at any other times. 
But there are many other components of the mixed state with $ \epsilon \ne \epsilon_{0}$. 
Any of them is available for the next detection, say, at time $t= t_{2} \ne t_{1}$. 
This is why for a mixed single-particle state the second-order correlation function is not vanishing, $G_{N=1} ^{(2)} \ne 0$. 
Obviously, all higher-order correlation functions are also not vanishing. 

The ability of a (mixed) single-particle state to demonstrate the second-order coherence can be verified (or refuted) experimentally. 
In particular, the function $G ^{(2)}$ with pairwise equal arguments is directly accessible through the cross-correlation noise measurement.

\section{  $G ^{(2)}$ and the cross-correlation noise}
\label{sec3}

\begin{figure}[b]
\includegraphics[width=70mm, angle=0]{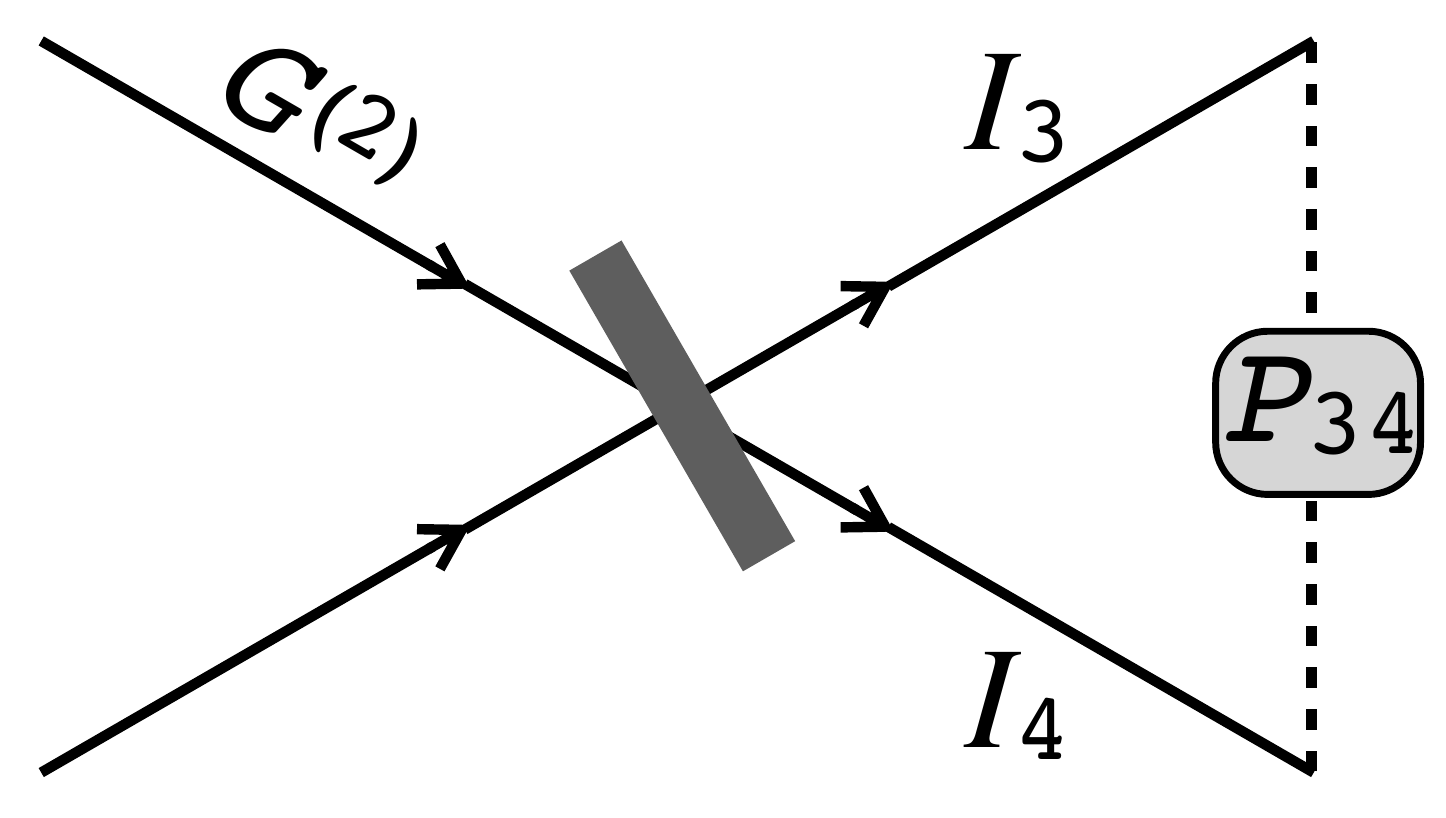}
\caption{
Scheme of an electron HBT interferometer, where the quantum state injected into one of input channels (marked by $G ^{(2)}_{}$) is transmitted through and reflected at the wave splitter (shown as a shaded thin rectangle). 
As a result the outgoing current $I_{3}$ and $I_{4}$ are generated. 
These currents, together with their cross-correlation function ${ P}_{34}$,  define the second-order correlation function of injected state, $G ^{(2)}_{}$, according to Eq.~(\ref{n03}). 
The arrows show the direction of propagation of electrons.
}
\label{fig1}
\end{figure}

As it was pointed-out in Ref.~\onlinecite{Thibierge:2015up}, the second-order correlation function with pairwise equal arguments, $G ^{(2)}\left( t_{1}, t_{2}; t_{2}, t_{1} \right)$, is directly related to the  cross-correlation symmetrized noise\cite{Buttiker:1990tn,Samuelsson:2004uv,Neder:2007jl,McClure:2007gl}. 
More precisely, it is related to the currents, $I_{3}\left( t_{1} \right)$ and $I_{4}\left( t_{2} \right)$, and their correlation function, ${ P}_{34}\left( t_{1}, t_{2} \right)$, which are measured at the outputs of an electronic interferometer \cite{Henny:1999tb,Oliver:1999ws,Oberholzer:2000wx}, analogous to the Hanbury Brown and Twiss (HBT) interferometer \cite{HanburyBrown:1956bi} known in optics, see Fig.~\ref{fig1}. 
The source of electrons is placed in one of the inputs. 
The temperature of both input channels $1$ and $2$, with and without an electron source, should be the same, $ \theta_{1} = \theta_{2} \equiv \theta$. 
Since the source injects particles periodically with period $ {\cal T} _{0}$, the resulting currents are periodic functions of time, $ I_{ \alpha}\left(  t \right) = I_{ \alpha}\left(t +  {\cal T} _{0} \right)$, $ \alpha = 3, 4$. 
 
Since the measurement of a time-resolved noise is challenging, below I focus on a frequency-resolved noise\cite{Buttiker:1992vn,Liu:1994fy,Lesovik:1994tq,Pedersen:1998uc,Salo:2006ie,Braggio:2016ki,Dittmann:2018ve}, which was measured more than once, see, e.g., Refs.~\onlinecite{{Schoelkopf:1997cc},DiCarlo:2006ju,ZakkaBajjani:2007gg} and also Ref.~\onlinecite{Mahe:2010cp}, where,  as I already mentioned, a frequency-resolved noise was used for validation of the single-electron injection regime. 

Let us introduce the following Fourier transform, 

\begin{eqnarray}
 G ^{(2)}_{ \ell}\left( \omega  \right) = 
 \int\limits _{0}^{ {\cal T} _{0} } dt e^{i \Omega \ell t}   
\int\limits _{- \infty}^{ \infty } d \tau e^{i \omega \tau}  
 G ^{(2)}\left( t + \tau, t; t, t + \tau \right) ,
\label{n01}
\end{eqnarray}
\ \\ \noindent
where $ \Omega = 2 \pi/ {\cal T} _{0}$ and $ \ell$ is an integer. 
Then, $ G ^{(2)}_{ \ell}\left( \omega  \right)$ is expressed in terms of the finite-frequency cross-correlation noise power, $ {\cal P}_{ 34,\ell}\left( \omega \right)$ and outgoing currents $I_{3}$ and $I_{4}$, as follows, (see the Appendix~\ref{ap01} for the precise definition of $ {\cal P}_{ 34,\ell}\left( \omega \right)$, Eqs.~(\ref{a01}) and (\ref{a02}), and for the corresponding derivation within the Floquet scattering matrix approach)

\begin{eqnarray}
v_{ \mu}^{2} G ^{(2)}_{ \ell}\left( \omega  \right) = 
\frac{{\cal P}_{34, \ell}(\omega) }{ e^{2} RT/ {\cal T} _{0} } 
\label{n02} \\
+
\frac{1 }{ e^{2} RT }
\int\limits _{0}^{ {\cal T} _{0} } e^{i \Omega \ell t} dt  
\int\limits _{- \infty}^{ \infty } d \tau e^{i \omega \tau}  
I_{3}\left( t + \tau \right) I_{4}\left( t \right) . 
\nonumber 
\end{eqnarray}
\ \\ \noindent  
Here $T$ and $R = 1-T$ are the transmission and reflection probabilities of a wave splitter of an electron HBT interferometer, $v_{ \mu}$ is the Fermi velocity of electrons in a waveguide, $e$ is an electron charge.  

In the special case, when the excitations produced during different periods do not overlap, the equation (\ref{n02}) can be simplified. 
We take into account explicitly the fact that the current is periodic, set $ {\cal T} _{0} \to \infty$, and introduce a continuous frequency $ \omega_{ \ell} = \ell \Omega$ instead of the series of discrete frequencies $ \ell \Omega$, see Appendix~\ref{ftg1} for details. 
Then the equation (\ref{n02}) becomes,

\begin{eqnarray}
v_{ \mu}^{2} G ^{(2)}_{ \ell}\left( \omega  \right) &=& 
\frac{{\cal P}_{34, \ell}(\omega) }{ e^{2} RT/ {\cal T} _{0}  } + \frac{ I_{3}\left( \omega \right) I_{4}\left( \omega_{ \ell} - \omega \right) }{ e^{2} RT/ {\cal T} _{0}^{2}} .
\label{n03}
\end{eqnarray}
\ \\ \noindent 
This equation
resembles Eq.~(22) of Ref.~\onlinecite{Moskalets:2014ea}, where the two-energy distribution function was related to the zero-frequency noise power and DC currents in the circuit with two energy filters, quantum dots each with one working resonant quantum level. 
 
Below, I will use Eq.~(\ref{n03}) and address the temperature dependence of $G ^{(2)}_{ \ell}\left(  \omega \right)$ for electrons injected   by some particular source, namely the source of levitons \cite{Dubois:2013ul}, that is capable of generating single- as well as multi-particle excitations \cite{Dubois:2013fs,Gaury:2014jz,Hofer:2014jb,Belzig:2016jza,Suzuki:2017er,Glattli:2017vp}.

\section{Example: The source of levitons}
\label{sec4}

The sequence of the Lorentzian voltage pulses,  

\begin{eqnarray}
eV(t) = N\sum\limits_{m=-\infty}^{\infty} \frac{2 \hbar  \Gamma _{\tau} }{\left( t - m {\cal T} _{0} \right)^{2} +  \Gamma _{\tau}^{2} }.
\label{L01}
\end{eqnarray}
\ \\ \noindent
applied to a metallic contact, generates the stream of excitations with charge $eN$ each in a ballistic channel attached to the contact. \cite{Levitov:1996,Ivanov:1997,Keeling:2006}
Here $  \Gamma _{\tau}$ is the half-width of a voltage pulse. 
These excitations are named $N$-electron levitons or $N$-levitons \cite{Ronetti:2017vd}.

\subsection{Correlation functions}

In the regime, when the period is much larger then the width of a voltage pulse,  $ {\cal T} _{0} \gg  \Gamma _{\tau}$, the excitations created at different periods do not overlap. 
Then, we can restrict ourselves to a single period only, say, $m=0$, and send  $ {\cal T} _{0} \to \infty$ in the integrals we need to evaluate. 
In this case the first-order correlation function of excitations injected by the source of levitons is represented as follows, 

\begin{eqnarray}
v_{ \mu} G^{(1)}_{  }( t_{1};t_{2} ^{}) =  
\int\limits_{ }^{ } d \epsilon^{} 
p_{ \theta} \left( \epsilon \right)
\sum\limits_{j=1}^{N}  
\Psi^{*}_{j, \epsilon}(t_{1}) \Psi_{j, \epsilon}(t_{2}^{}) .
\label{L02} 
\end{eqnarray}
\ \\ \noindent
Here $\Psi_{j, \epsilon}(t) =  e^{-i t \frac{ \mu + \epsilon }{ \hbar} } \psi_{j}\left(  t \right)$ is the wave function of the $j$th particles  comprising an $N$-electron leviton ($j = 1, \dots, N$). 
The corresponding envelope function is the following, \cite{Grenier:2013,Moskalets:2015vr,glattli2016method} 

\begin{eqnarray}
\psi_{j}\left(  t \right) &=&
\sqrt{\frac{\Gamma _{\tau}}{  \pi  } }
\frac{ 1 }{  t  - i \Gamma _{\tau} }
\left(  \frac{ t + i \Gamma _{\tau}  }{  t  - i \Gamma _{\tau}  } \right) ^{j-1} .
\label{L03} 
\end{eqnarray}
\ \\ \noindent
Using the fact that the envelope wave functions, $ \psi_{j}$, are independent of energy, we can integrate $ \epsilon$ out in Eq.~(\ref{L02}) and get,

\begin{eqnarray}
v_{ \mu} G^{(1)}_{  }( t_{1};t_{2} ^{}) =  
\eta\left( \frac{ t_{1} - t_{2} }{ \tau_{ \theta} } \right)
\sum\limits_{j=1}^{N}  
\psi^{*}_{j}(t_{1}) \psi_{j}(t_{2}^{}) ,
\label{L04} 
\end{eqnarray}
\ \\ \noindent
where $ \eta(x) = x/\sinh(x)$ and the thermal coherence time is $ \tau_{ \theta} = \hbar/( \pi k_{B} \theta)$.  

Substituting the above equation into  Eq.~(\ref{gn01}), one can   calculate the second-order correlations function.  
For $t_{1} = t_{4}$ and $t_{2} = t_{3}$ we have,

\begin{eqnarray}
v_{ \mu}^{2}G ^{(2)}\left( t_{1},t_{2};t_{2},t_{1} \right) =
\sum\limits_{j=1}^{N}  
\sum\limits_{k=1}^{N}  
\Bigg\{
\left | \psi^{}_{j}(t_{1})  \right |^{2}
\left | \psi^{}_{k}(t_{2})  \right |^{2}
\nonumber \\
\label{L05} \\
- 
\eta^{2}\left( \frac{ t_{1} - t_{2} }{ \tau_{ \theta} } \right)
\psi^{*}_{j}(t_{1}) 
\psi^{}_{j}(t_{2}) 
\psi^{*}_{k}(t_{2}) 
\psi^{}_{k}(t_{1}) 
\Bigg\}. 
\nonumber 
\end{eqnarray}

Now I will analyze the above equation in two cases, $N=1$ and $N=2$. 

\subsection{A single-electron leviton, $N=1$}

For a single-particle leviton, $N=1$, the function  $G ^{(2)}$ becomes,

\begin{eqnarray}
v_{ \mu}^{2} G_{N=1} ^{(2)}\left( t_{1},t_{2};t_{2},t_{1} \right) = 
\frac{  \Gamma _{\tau}^{2} }{ \pi^{2} } 
\frac{ 1 - \eta^{2}\left( \frac{ t_{1} - t_{2} }{ \tau_{ \theta} } \right) }{ \left(  t_{1}^{2} +  \Gamma _{\tau}^{2} \right) \left(  t_{2}^{2} +  \Gamma _{\tau}^{2} \right) } .
\label{L06}
\end{eqnarray}
\ \\ \noindent
From this equation, we can conclude the following. 
First, when the time difference is smaller then the thermal coherence time, $\left | t_{1} - t_{2} \right | \ll \tau_{ \theta}$, the function $ \eta = 1$, and the second order correlation function vanishes, $G_{N=1} ^{(2)} =0$. 
This fact is a manifestation of a single-particle nature of a quantum state in question. 

Second, at larger time difference, $\left | t_{1} - t_{2} \right | \gg \tau_{ \theta}$, the  function $ \eta = 0$, and the second order correlation function is factorized into the product of two terms, each of which depends only on one time, $G_{N=1} ^{(2)}\left( t_{1},t_{2};t_{2},t_{1} \right) = \left | \psi_{1}\left( t_{1} \right) \right |^{2} \left | \psi_{1}\left( t_{2} \right) \right |^{2}$. 
Namely, the two-particle detection probability becomes the product of two statistically independent single-particle detection probabilities. 
Such a property is expected for a classical rather than a quantum state. 
Nevertheless, the state of a leviton remains quantum and respects the Pauli exclusion principle, which requires that the function $G ^{(2)}$ strictly vanishes  at equal  times (at any temperature),  $G_{N=1} ^{(2)}\left( t,t;t,t \right) =0$.

\subsubsection{The frequency representation}

Let us perform the Fourier transformation defined in Eq.~(\ref{n01}) on the  function $G ^{(2)}$ of a single leviton, Eq.~(\ref{L06}). 
Using the fact that $ {\cal T} _{0} \gg  \Gamma _{\tau}$, we get,

\begin{eqnarray}
v_{ \mu}^{2} G_{N=1, \ell} ^{(2)}\left(  \omega \right) = 
e^{- \left | \omega \right |  \Gamma _{\tau}}
e^{- \left | \omega - \omega_{\ell} \right |  \Gamma _{\tau}}
-
e^{-  \left | \omega_{ \ell} \right |  \Gamma _{\tau}}
\nonumber \\
\times
\frac{   \Gamma _{\tau} }{ \pi }
\int\limits _{- \infty}^{ \infty } d \tau 
\frac{ \eta^{2}\left( \frac{ \tau }{ \tau_{ \theta} } \right) }{ \tau^{2} + 4  \Gamma _{\tau}^{2} } 
\bigg\{
\cos\left( \omega \tau  \right)
 + 
\cos\left( \left [ \omega - \omega_{ \ell} \right] \tau  \right)
\nonumber \\
+ \frac{ 2  \Gamma _{\tau} }{ \tau }  {\rm sgn}( \omega_{ \ell})  
\left[ 
\sin\left( \omega \tau  \right)
- 
\sin\left( \left [ \omega  - \omega_{ \ell} \right] \tau \right)
\right] 
\bigg\} .
\nonumber \\
\label{Lom01} 
\end{eqnarray}
As I already mentioned, this function is experimentally accessible through the  finite-frequency noise measurement, see Eq.~(\ref{n03}). 

\begin{figure}[t]
\includegraphics[width=85mm, angle=0]{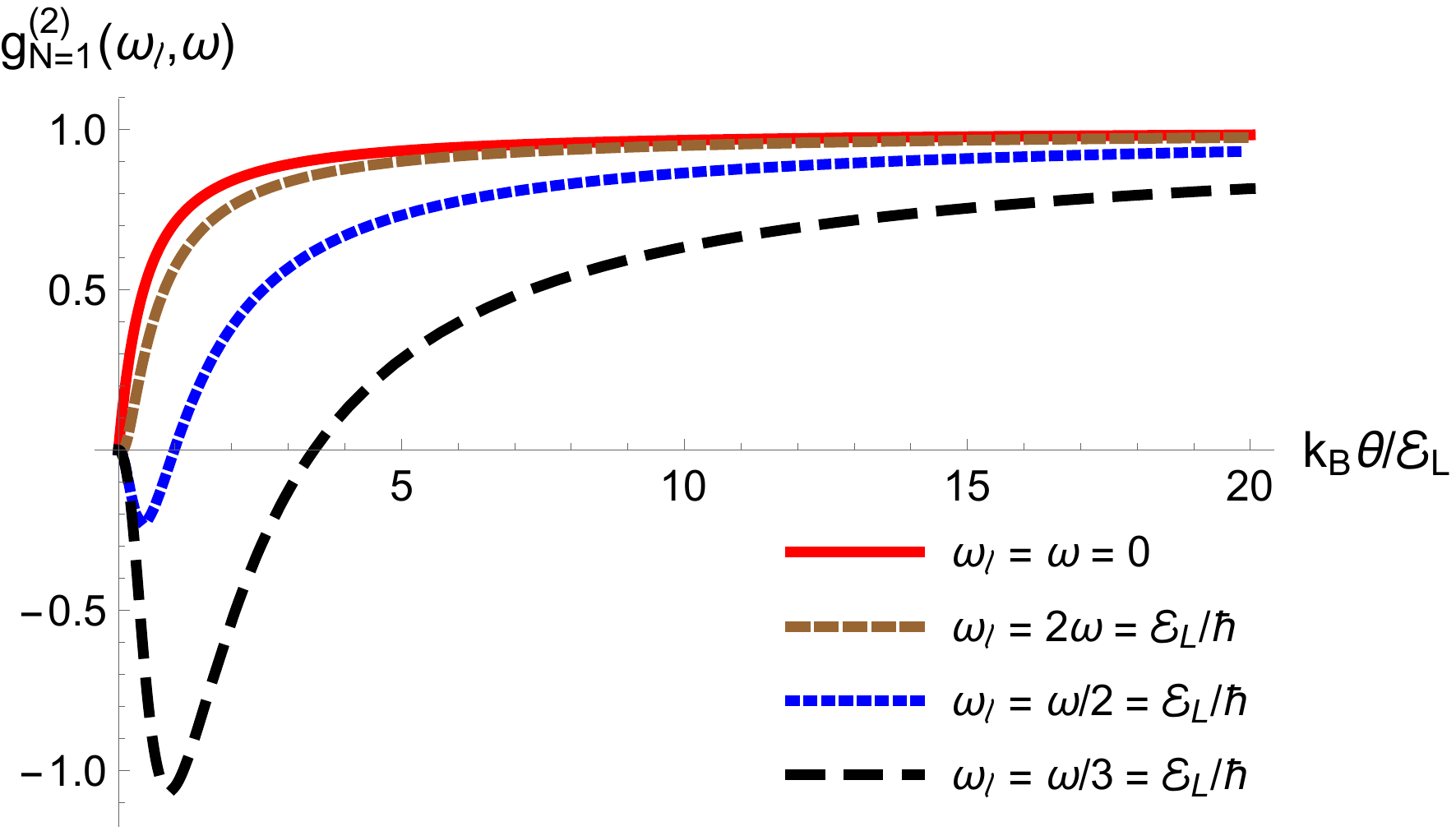}
\caption{
The second-order correlation function of a single leviton, Eq.~(\ref{Lom01}), normalized to its high-temperature asymptotics, $g_{N=1} ^{(2)}\left( \omega_{ \ell},  \omega \right)  = v_{ \mu}^{2} G_{N=1, \ell} ^{(2)}\left(  \omega \right)/\left( e^{- \left | \omega \right |  \Gamma _{\tau}} e^{- \left | \omega - \omega_{\ell} \right |  \Gamma _{\tau}}  \right) $, is given as a function of temperature.  
The temperature, $k_{B} \theta$, and frequencies, $ \hbar \omega$ and $ \hbar  \omega_{ \ell}$, are given in units of the energy of a leviton, $ {\cal E}_{L} = \hbar / \left( 2  \Gamma _{\tau} \right)$. 
}
\label{fig2}
\end{figure}

In Fig.~\ref{fig2} I show $G_{N=1, \ell} ^{(2)}\left(  \omega \right)$, Eq.~(\ref{Lom01}), as a function of temperature for several fixed frequencies. 
My aim is to show that the function $G ^{(2)}_{N=1}$ is capable of demonstrating a crossover from a single-particle behaviour at zero temperature to a multi-particle-like behaviour at nonzero temperatures.    
Indeed, at zero temperature $G_{N=1, \ell} ^{(2)}\left(  \omega \right)=0$ for any frequencies, demonstrating that the state in question is a true single-particle state. 
At nonzero temperatures, the function $G ^{(2)}$ becomes different from zero,  indicating that the state of a single leviton demonstrates rather multi-particle behaviour. 
At high temperatures, when  the thermal coherence time becomes  smaller than the width of a voltage pulse, $\tau_{ \theta} \ll  \Gamma _{\tau}$, the second-order correlation function achieves its high-temperature asymptotic behaviour, $v_{ \mu}^{2}\lim\limits_{ \theta\to \infty} G_{N=1, \ell} ^{(2)}\left(  \omega \right)=e^{- \left | \omega \right |  \Gamma _{\tau}} e^{- \left | \omega - \omega_{\ell} \right |  \Gamma _{\tau}}$. 
This product form is characteristic of a completely classical state. 

Note, that the growth of $G ^{(2)}_{N=1}$ with temperature is manifested in the reduction of single-particle shot noise\cite{Dubois:2013fs,Moskalets:2015ub,Moskalets:2017dy}. 
The shot noise decrease with increasing temperature was reported in Refs.~\onlinecite{Bocquillon:2012if,{Bocquillon:2013fp},Glattli:2016wl,Glattli:2016tr}. 

The temperature dependence of the function $G ^{(2)}$ for a multi-electron state is remarkably different. 
Namely, its zero temperature limit is not universal. 
On contrary, such a limit depends strongly on frequency.
To illustrate this statement, let us consider the case of a $2$-electron leviton.

\subsection{A two-electron leviton, $N=2$}

\begin{figure}[t]
\includegraphics[width=85mm, angle=0]{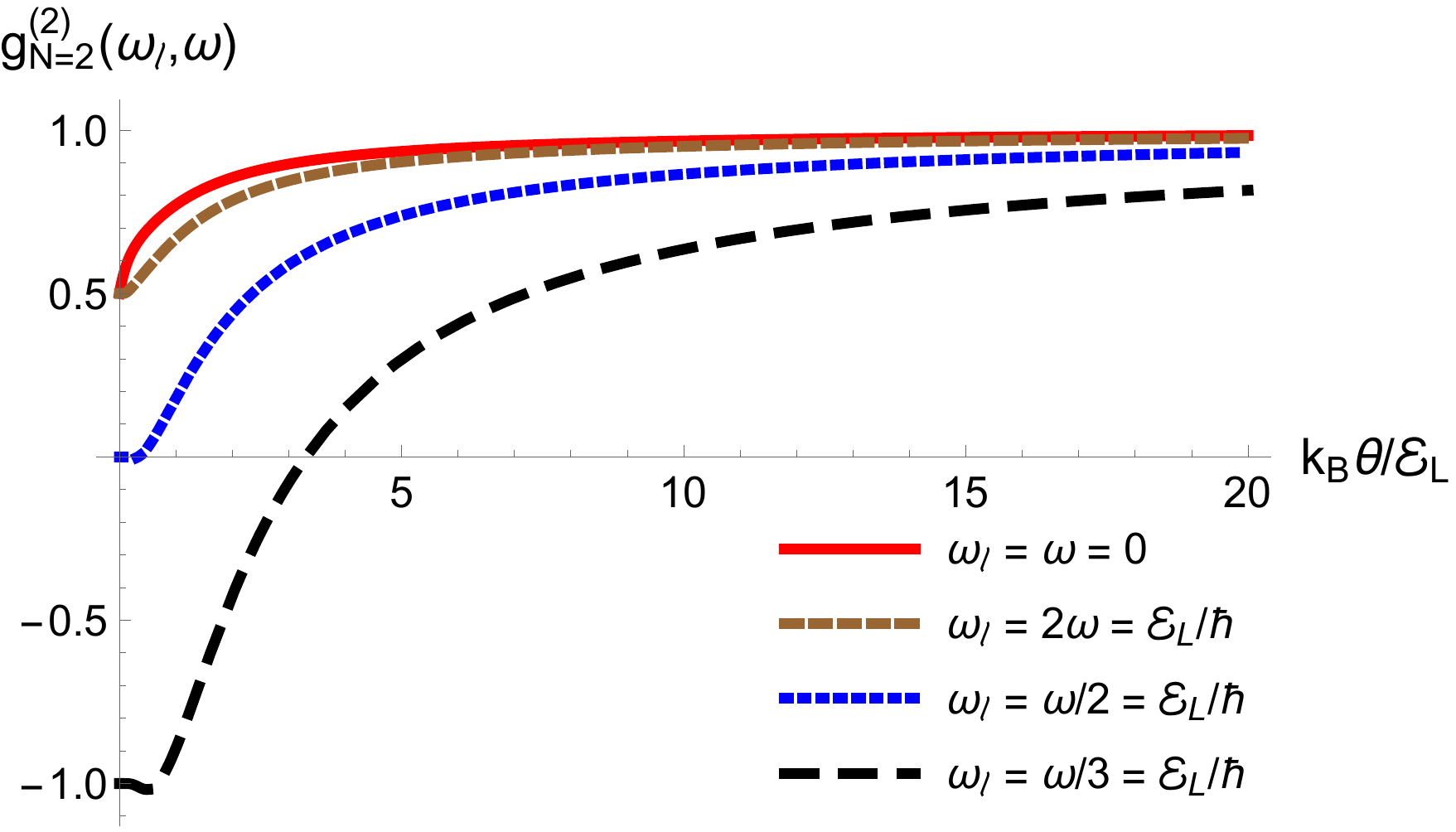}
\caption{
The second-order correlation function of a $2$-electron leviton, Eq.~(\ref{b04}), normalized to its high-temperature asymptotics, $g_{N=2} ^{(2)}\left( \omega_{ \ell},  \omega \right)  = v_{ \mu}^{2} G_{N=2, \ell} ^{(2)}\left(  \omega \right)/\left(4 e^{- \left | \omega \right |  \Gamma _{\tau}} e^{- \left | \omega - \omega_{\ell} \right |  \Gamma _{\tau}}  \right) $, is given as a function of temperature. 
The temperature, $k_{B} \theta$, and frequencies, $ \hbar \omega$ and $ \hbar  \omega_{ \ell}$, are given in units of the energy of a leviton, $ {\cal E}_{L} = \hbar / \left( 2  \Gamma _{\tau} \right)$. 
}
\label{fig3}
\end{figure}

The correlation function of a $2$-electron leviton,  $G_{N=2} ^{(2)}$, is given in Eq.~(\ref{L02}) with $N=2$. 
The corresponding wave functions are presented in Eq.~(\ref{L03}). 
After performing the Fourier transformation according to Eq.~(\ref{n01}), we obtain $G_{N=2, \ell} ^{(2)}\left(  \omega \right)$, see Eq.~(\ref{b04}),  which is shown in Fig.~\ref{fig3} as a function of temperature for several fixed frequencies.

We can see, that at zero temperature, the magnitude of $G_{N=2, \ell} ^{(2)}\left(  \omega \right)$ is not zero, unlike the case of a plain leviton, $N=1$. 
Indeed, it strongly depends on the frequencies $ \omega_{ \ell}$ and $ \omega$. 
Such a nonuniversal frequency-dependent behaviour is characteristic of a multi-particle state, the two-particle state in the present case. 

Interestingly, the high-temperature asymptotics of the second-order correlation function of a multi-electron leviton is universal: It is determined by the corresponding asymptotics of the function $G ^{(2)}$ of a single-electron leviton, 

\begin{eqnarray}
\lim\limits_{ \theta\to \infty} G_{N=N_{0}} ^{(2)} = N_{0}^{2} \lim\limits_{ \theta\to \infty} G_{N=1} ^{(2)} .
\label{L07}
\end{eqnarray}
\ \\ \noindent
This is a manifestation of the high-temperature fusion effect (when the multi-electron system behaves like one particle of the total charge) discussed in Ref.~\onlinecite{Moskalets:2018ch}.   

Note, that the low-temperature regime,  for which Figs.~\ref{fig2} and \ref{fig3} show different behaviour, is achievable in present day experiment. 
So, in Ref.~\onlinecite{Glattli:2016tr} the voltage pulses with width $2 \Gamma _{\tau} = 75$~ps were user to generate levitons with energy $ {\cal E}_{L} \approx 320$~mK.  
The experimental data on shot noise  were reported for the temperature range from $ \theta_{1} = 40$~mK to $ \theta_{2} = 138$~mK. 
Correspondingly,  the ratio $ \theta / {\cal E}_{L}$ is changed from  $ \theta_{1} / {\cal E}_{L} \approx 0.125$ to $ \theta_{2} / {\cal E}_{L} \approx 0.43$.   
From  Figs.~\ref{fig2} and \ref{fig3} we see that for these parameters the function $G ^{(2)}_{}$ allows one to uniquely distinguish single-particle and multi-particle states: The second-order correlation function, $G ^{(2)}_{}$, is almost constant in the case of a $2$-electron leviton, while it decreases rapidly to zero with decreasing temperature in the case of a single-electron leviton.

\section{Conclusion}
\label{sec5}

I have discussed the effect of temperature on the second-order correlation function of electrons,  $G ^{(2)}$, which are injected by an on-demand source on top of the Fermi sea in conductors. 

The second-order correlation function is a universal tool that is able to distinguish between single- and multi-particle injection regime of an electron source. 
The function $G ^{(2)}$  is accessible via the cross-correlation electrical noise measurement at the exit of an electron Hanbury Brown and Twiss interferometer.    

At zero temperature, the function  $G ^{(2)}$ is vanishing in the case of a single-electron injection and does not vanish in the case of multi-particle injection.  
In contrast, at nonzero temperatures, the function  $G ^{(2)}$ does not vanish even in the case of a single-electron injection. 
The reason is that at nonzero temperatures, the single-particle quantum state  is a mixed state that demonstrates some degree of second-order coherence, which is quantified by  $G ^{(2)} \ne 0$. 
Therefore, the existence of this single-particle contribution has to be taken into account, when the second-order correlation function is used for the verification of a single-particle injection into conductors at nonzero temperatures.

\acknowledgments

I appreciate the warm hospitality of the Department of Applied Physics, Aalto University, Finland, where this project was started. 
I am grateful to Christian Flindt and Pablo Burset for numerous discussions.

\appendix

\section{Electron versus electrical correlation functions}
\label{ap01}

The setup of interest consists of an electron wave splitter, a quantum point contact \cite{VanWees:1988vf,{Wharam:1988vi}}, with two incoming, $ \alpha=1,2$, and two outgoing, $ \beta = 3,4$, one-dimensional wave guides, which are connected to respective metallic leads with the same temperature $ \theta$ and chemical potential $ \mu$, see Fig.~\ref{fig1}. 
The transmission $T$ and reflection $R=1-T$ probabilities of a wave splitter are energy independent (within the energy range relevant to our problem). 

One of the incoming waveguides, say $ \alpha=1$, is connected to an electron source that periodically emits particles. 
The period is denoted as $ {\cal T} _{0}$. 
Within the scattering approach, the periodically working source is characterized by the Floquet scattering matrix $S_{F}$ dependent on two energies, $E$ and $E_{n} = E + n \hbar \Omega $ with $n$ being integer and the frequency $ \Omega = 2 \pi/ {\cal T} _{0}$ is dictated by the periodicity. 
By the virtue of definition, $S_{F}\left( E_{n}, E \right)$ is a quantum-mechanical amplitude for the process, when an electron with energy $E$  passing by a source changes its energy to $E_{n}$. \cite{Moskalets:2002hu}

My aim here is to prove Eq.~(\ref{n02}). 
For this, I use Eq.~(\ref{gn01}) and write, 

\begin{eqnarray}
G ^{(2)}_{}\left(  t_{1},t_{2}; t_{2}, t_{1} \right) &=& 
- \left | G ^{(1)}_{}\left(  t_{1};t_{2} \right) \right |^{2}
\nonumber \\
\label{a01-2} \\
&&
+
G ^{(1)}_{}\left(  t_{1};t_{1} \right) G ^{(1)}_{}\left(  t_{2};t_{2} \right) 
 .
\nonumber 
\end{eqnarray}
\noindent \\ 
The first-order correlation function at coincident times defines an electrical  current generated by the source, $I_{}(t) = e v_{ \mu}G ^{(1)}\left( t;t \right)$.\cite{Moskalets:2016va} 
The outgoing currents are expressed in terms of the incoming current $I_{}(t)$, as follows,  $I_{3}(t) = R I_{}(t)$ and $I_{4}(t) = T I_{}(t)$. 
The currents are periodic in time, $I(t) = I(t + {\cal T} _{0})$.  
Therefore, after the Fourier transform defined in Eq.~(\ref{n01}), the second line   of Eq.~(\ref{a01-2}) reproduces the second line of Eq.~(\ref{n02}). 

To prove that the first term on the right hand side of Eq.~(\ref{a01-2}) leads to the first term on the right hand side of Eq.~(\ref{n02}), I express  both the cross-correlation electrical noise power of the outgoing currents, ${\cal P}_{34}$, and the excess first-order correlation function of electrons injected by the source, $G ^{(1)}_{}$, in terms of the Floquet scattering matrix of the source, and then relate them to each other.

\subsection{Frequency-dependent electrical noise}

The symmetrized cross-correlation function of  electrical currents $I_{3}$ and $I_{4}$ flowing out of the wave splitter, see Fig.~\ref{fig1}, is defined as follows, \cite{Buttiker:1992vr}

\begin{eqnarray}
{ P}_{ 34}\left(  t_{1}. t_{2} \right) &=& 
\frac{1 }{2 } 
\left\langle \Delta \hat I_{ 3}(t) \Delta \hat I_{ 4}(t + \tau)  + \Delta \hat I_{ 4}(t + \tau)  \Delta \hat I_{ 3}(t)  \right\rangle ,
\nonumber \\
\label{a01-1} 
\end{eqnarray}
\ \\ \noindent
where $\Delta \hat I_{ \alpha}(t) =  \hat I_{ \alpha}(t) - \left\langle  \hat I_{ \alpha}(t) \right\rangle$, $ \alpha = 3,4$, is an operator of current fluctuations in second quantization. 
The angle brackets  $\left\langle \dots  \right\rangle $ denote a quantum-statistical average over the equilibrium state of an incoming single-mode channel not affected by the electron source.  
Such an equilibrium state is the Fermi sea with a temperature $ \theta$ and a chemical potential $ \mu$. 
The other incoming channel of the wave splitter is in the same  equilibrium state, with the same temperature $ \theta$ and the same chemical potential $ \mu$. 
 
For convenience, let us analyze this quantity in frequency representation.  
In the general non-stationary case, the current correlation function $ { P} _{34}$ depends on two frequencies. 
In the case with periodic driving, the Floquet scattering theory calculations give for our setup,\cite{Moskalets:2007dl,Moskalets2013}    

\begin{eqnarray}
{ P}_{34}(\omega,\omega^{\prime}) &=& 2\pi \sum\limits_{ \ell=-\infty}^{\infty}
 \delta(\omega + \omega^{\prime} - \ell\Omega) 
{\cal P}_{34, \ell}(\omega),
\label{a01}
\end{eqnarray}
\ \\ \noindent  
where the noise power ${\cal P}_{34, \ell}(\omega)$ is expressed in terms of the Floquet scattering matrix elements of an electron source, $S_{F}$, as follows,

\begin{eqnarray}
{\cal P}_{34, \ell}(\omega) = - RT\frac{e^2}{h}
\int\limits_{}^{}dE
\sum\limits_{n}\sum\limits_{m}\sum\limits_{q}
f_{}(E_q + \hbar\omega) 
f_{}(E_{}) 
\nonumber \\
\label{a02} \\
\left\{ 
\delta_{n0}\delta_{m+\ell,0} -
S_{F}^{*}(E_{n},E_{})  
S_{F}(E_{m+ \ell}, E_{})
\right\}
\Big\{ 
\delta_{nq}\delta_{mq}
\nonumber \\
\nonumber \\
- 
S_{F}^{*}(E_{m}+\hbar\omega,E_{q}+\hbar\omega)
S_{F}(E_{n}+\hbar\omega, E_q+\hbar\omega) 
\Big\}
.
\nonumber 
\end{eqnarray}
\ \\ \noindent
Here $n,m,q$ are integers, $ \delta_{nm}$ is the Kronecker delta, and  $f\left( E \right) = \left(1 + e^{- \frac{E - \mu }{ k_{B} \theta }}  \right)^{-1}$ is the Fermi distribution function for electrons in a metallic contact with temperature $ \theta$ and chemical potential $ \mu$, $k_{B}$ is the Boltzmann constant.  

Easy to see that in equilibrium, that is, when the source is turned off and,  accordingly, $S_{F}\left( E_{n},E_{m} \right) = \delta_{nm}$, the noise power is zero, ${\cal P}_{34, \ell}(\omega) = 0$. 
So, in our setup, where the contacts $ \beta = 3$ and $ \beta = 4$ have no a direct connection between themselves, only the partition noise of  injected particles, but not the quantum noise \cite{Gardiner2000}, contributes to the measured noise.  

For convenience of the subsequent comparison with the electron correlation function, I represent the noise power as the sum of four terms, ${\cal P}_{34, \ell}(\omega) = - RT\frac{e^2}{h} \sum_{r=1}^{4} B_{r}$, with  

\begin{subequations}
\label{a03}

\begin{eqnarray}
B_{1}^{} = 
\int\limits_{}^{}dE
\sum\limits_{n}\sum\limits_{m}
\sum\limits_{q}
f_{}(E_{q} + \hbar\omega) f_{}(E_{}) 
\quad\quad
\label{a03a} \\
S_{F}^{*}(E_{n},E_{})  
S_{F}(E_{m+ \ell}, E_{})
S_{F}^{*}(E_{m}+\hbar\omega,E_{q}+\hbar\omega)
\nonumber \\
S_{F}(E_{n}+\hbar\omega, E_q+\hbar\omega) 
 ,
\nonumber 
\end{eqnarray}

\begin{eqnarray}
 \quad \
B_{2}^{} = -
\int\limits_{}^{}dE
\sum\limits_{q}
f_{}(E_{q} + \hbar\omega) f_{}(E_{}) 
\quad\quad
\label{a03b} \\
S_{F}^{*}(E_{q}, E_{}) 
S_{F}(E_{q+ \ell}, E_{}) 
 ,
\nonumber 
\end{eqnarray}

\begin{eqnarray}
B_{3}^{} = -
\int\limits_{}^{}dE
\sum\limits_{q}
f_{}(E_{q} + \hbar\omega) f_{}(E_{}) 
\quad\quad
\label{a03c} \\
S_{F}^{*}(E_{- \ell}+\hbar\omega, E_q+\hbar\omega) 
S_{F}(E_{}+\hbar\omega, E_q+\hbar\omega) ,
\nonumber 
\end{eqnarray}

\begin{eqnarray}
B_{4}^{} &=&  \delta_{ \ell 0}
\int\limits_{}^{}dE
f_{}(E + \hbar\omega) f_{}(E_{})  .
\quad\quad
\label{a03d} 
\end{eqnarray}

\end{subequations}

Now let us turn to the quantum description of the excitations injected by the source.

\subsection{Electron correlation function}

As I discussed in Sec.~\ref{sec2}, the first-order correlation function for these excitations,  $G ^{(1)}$, is defined as the difference of the electron correlation functions  with the source   switched on and off, $  G ^{(1)}\left( t; t ^{\prime} \right) = \left\langle \hat \Psi^{ \dag}\left( t \right) \hat \Psi\left(  t ^{\prime} \right) \right\rangle_{on} - \left\langle \hat \Psi^{ \dag}\left( t \right) \hat \Psi \left(  t ^{\prime} \right) \right\rangle_{off}$. 
Here $ \hat \Psi\left( t \right)$ is a single-particle electron field operator in second quantization at time $t$ just downstream the source. 
The angle brackets, $ \left\langle \dots \right\rangle$, denote the quantum-statistical average over the state of the Fermi sea of electrons approaching the source from the metallic contact $ \alpha=1$ being in equilibrium. 

I adopt the wide band approximation and suppose that in the waveguides the electron spectrum can be linearized. 
The Floquet scattering theory expresses $G ^{(1)}$ in terms of $S_{F}$ as follows, \cite{Moskalets:2015ub}

\begin{eqnarray}
G^{(1)}_{ }(t;t ^{\prime}) = 
\frac{ 1 }{ h v_{ \mu} }
\int dE f_{ }\left( E \right) 
e^{ \frac{ i }{ \hbar  } E   \left( t - t ^{\prime} \right)  } 
\quad\quad
\label{a04} \\
\left\{ 
\sum_{n,m=-\infty}^{\infty} 
e ^{i \Omega \left( n t - m t ^{\prime} \right)}
S_{F}^{*}\left(E_{n},E  \right)
S_{F}\left(E_{m},E  \right)
- 1 \right\} .
\nonumber 
\end{eqnarray}

Note, that the possibility to linearize an electron spectrum is important to perform linear optic-like manipulations with electrons. 
For instance, if we need to calculate  $G ^{(1)}$ at another place, say, at a distance $x$ from the source downstream, we simply replace $t \to t - v_{ \mu} x$ with $v_{ \mu}$ being the Fermi velocity. 

The wide band approximation is also crucial to get simple relations between electrical and quantum-mechanical quantities.

\subsection{The connection between the two}

Now let us demonstrate that the electrical noise power and the electron correlation function squared are related by the following Fourier transformation,

\begin{eqnarray}
\frac{{\cal P}_{34, \ell}(\omega) }{ {\cal P}_{0}  } = -
\int\limits _{0}^{ {\cal T} _{0} } e^{i \Omega \ell t} dt  
\int\limits _{- \infty}^{ \infty } d \tau e^{i \omega \tau} \left | v_{ \mu} 
G^{(1)}_{ }(t_{}+ \tau;t_{} ) \right | ^{2} .
\nonumber \\
\label{a05}
\end{eqnarray}
\ \\ \noindent
Here $ {\cal P} _{0} = e^{2} RT / {\cal T} _{0}$ is the circuit constant. 
This constant is the shot noise caused by the scattering of single electrons being in the pure state on a wave splitter at a rate of one particle per period $ {\cal T} _{0}$. \cite{Blanter:2000wi}

The relation analogous to Eq.~(\ref{a05}) but for the zero-frequency noise power was presented in Refs.~\onlinecite{Moskalets:2015ub,Moskalets:2017dy}.

To prove Eq.~(\ref{a05}), first, let us represent the square of the correlation function, Eq.~(\ref{a04}), as the sum of four terms,

\begin{subequations}
\label{a06}

\begin{eqnarray}
\left | v_{ \mu} G^{(1)}_{ }(t_{};t ^{\prime}) \right | ^{2}   = \frac{ 1 }{ h^{2} } \sum\limits_{s=1}^{4} A_{s} (t_{};t ^{\prime}), 
\label{a06a}
\end{eqnarray}

\begin{eqnarray}
A_{1} (t_{};t ^{\prime}) = 
\int dE  f_{ }\left( E \right) 
e^{ \frac{ i }{ \hbar  } E   \left( t - t ^{\prime} \right)  } 
\int dE ^{\prime}  f_{ }\left( E ^{\prime} \right) 
\nonumber \\
\label{a06b} \\
e^{ \frac{ -i }{ \hbar  } E ^{\prime}  \left( t - t ^{\prime} \right)  }
\sum_{n,m}^{} 
e ^{i \Omega \left( n t - m t ^{\prime} \right)}
S_{F}^{*}\left(E_{n},E  \right)
S_{F}\left(E_{m},E  \right)
\nonumber \\
\sum_{j,k}^{} 
e ^{-i \Omega \left( j t - k t ^{\prime} \right)}
S_{F}^{}\left(E_{j} ^{\prime},E ^{\prime}  \right)
S_{F}^{*}\left(E_{k} ^{\prime},E ^{\prime}  \right) 
,
\nonumber 
\end{eqnarray}

\begin{eqnarray}
A_{2} (t_{};t ^{\prime}) = - 
\int dE  f_{ }\left( E \right) 
e^{ \frac{ i }{ \hbar  } E   \left( t - t ^{\prime} \right)  } 
\int dE ^{\prime}  f_{ }\left( E ^{\prime} \right) 
\nonumber \\
\label{a06c}\\
e^{ \frac{ -i }{ \hbar  } E ^{\prime}  \left( t - t ^{\prime} \right)  }
\sum_{n,m}^{} 
e ^{i \Omega \left( n t - m t ^{\prime} \right)}
S_{F}^{*}\left(E_{n},E  \right)
S_{F}\left(E_{m},E  \right)
,
\nonumber 
\end{eqnarray}

\begin{eqnarray}
A_{3} (t_{};t ^{\prime}) = -
\int dE  f_{ }\left( E \right) 
e^{ \frac{ i }{ \hbar  } E   \left( t - t ^{\prime} \right)  } 
\int dE ^{\prime}  f_{ }\left( E ^{\prime} \right) 
\nonumber \\
\label{a06d}\\
e^{ \frac{ -i }{ \hbar  } E ^{\prime}  \left( t - t ^{\prime} \right)  }
\sum_{j,k}^{} 
e ^{-i \Omega \left( j t - k t ^{\prime} \right)}
S_{F}^{}\left(E_{j} ^{\prime},E ^{\prime}  \right)
S_{F}^{*}\left(E_{k} ^{\prime},E ^{\prime}  \right)
,
\nonumber 
\end{eqnarray}

\begin{eqnarray}
A_{4} (t_{};t ^{\prime}) = 
\int dE  f_{ }\left( E \right) 
e^{ \frac{ i }{ \hbar  } E   \left( t - t ^{\prime} \right)  } 
\int dE ^{\prime}  f_{ }\left( E ^{\prime} \right) 
e^{ \frac{ -i }{ \hbar  } E ^{\prime}  \left( t - t ^{\prime} \right)  } .
\nonumber \\
\label{a06e}
\end{eqnarray}
\ \\ \noindent
\end{subequations}
As the next step, let us perform the following Fourier transformation,

\begin{subequations}
\label{a07}

\begin{eqnarray}
A_{s, \ell}\left( \omega \right) = 
\frac{ 1 }{ h }
\int _{0}^{ {\cal T} _{0} } \frac{dt  }{ {\cal T} _{0} }
 e^{i \Omega \ell t} 
\int \limits_{- \infty}^{ \infty } d \tau e^{i \omega \tau} A_{s}\left( t_{} + \tau, t_{}  \right) ,
\label{a07a}
\end{eqnarray}
\ \\ \noindent
and show that $A_{s, \ell}\left( \omega \right)$ is nothing but $B_{s}$, Eqs.~(\ref{a03}). 

Let us start with $A_{4}$,

\begin{eqnarray}
A_{4, \ell}\left( \omega \right) &=&
\frac{ 1 }{ h }
\int _{0}^{ {\cal T} _{0} } \frac{dt  }{ {\cal T} _{0} }  e^{i \Omega \ell t} 
\int\limits _{- \infty}^{ \infty } d \tau e^{i \omega \tau}
\label{a07b} \\
&&
\int dE  f_{ }\left( E \right) 
e^{ \frac{ i }{ \hbar  } E   \tau } 
\int dE ^{\prime}  f_{ }\left( E ^{\prime} \right) 
e^{ \frac{ -i }{ \hbar  } E ^{\prime} \tau  }
\nonumber \\
&=& 
\delta_{ \ell 0}
\int dE  f_{ }\left( E \right) \left( E+ \hbar \omega \right) . 
\nonumber 
\end{eqnarray}
\ \\ \noindent
This equation is exactly $B_{4}^{}$, Eq.~(\ref{a03d}). 

The Fourier transform of $A_{3}$, Eq.~(\ref{a06d}), gives us,

\begin{eqnarray}
A_{3, \ell}\left( \omega \right) = -
\frac{ 1 }{ h }
\int _{0}^{ {\cal T} _{0} } \frac{dt  }{ {\cal T} _{0} }  e^{i \Omega \ell t} 
\int\limits _{- \infty}^{ \infty } d \tau e^{i \omega \tau} 
\quad \
\label{a07c} \\
\int dE  f_{ }\left( E \right) 
e^{ \frac{ i }{ \hbar  } E   \tau  } 
\int dE ^{\prime}  f_{ }\left( E ^{\prime} \right) 
e^{ \frac{ -i }{ \hbar  } E ^{\prime}  \tau }
\nonumber \\
\sum_{j,k}^{} 
e ^{-i \Omega j \tau }
e ^{-i \Omega t \left( j - k \right)}
S_{F}^{}\left(E_{j} ^{\prime},E ^{\prime}  \right)
S_{F}^{*}\left(E_{k} ^{\prime},E ^{\prime}  \right)
\nonumber \\
= -
\int dE  
\sum_{j}^{} 
f_{ }\left( E \right) 
f_{ }\left( E_{-j} + \hbar \omega \right) 
\nonumber \\
S_{F}^{*}\left(E_{- \ell} + \hbar \omega,E_{-j} + \hbar \omega  \right) 
S_{F}^{}\left(E_{} + \hbar \omega,E_{-j} + \hbar \omega  \right) .
\nonumber 
\end{eqnarray}
\ \\ \noindent
After replacing $-j$ by $q$ we recognize the above equation as $B_{3}^{}$, Eq.~(\ref{a03c}). 

The next term is $A_{2}$, Eq.~(\ref{a06c}), 

\begin{eqnarray}
A_{2, \ell}\left( \omega \right) = -
\frac{ 1 }{ h }
\int _{0}^{ {\cal T} _{0} } \frac{dt  }{ {\cal T} _{0} } e^{i \Omega \ell t} 
\int\limits _{- \infty}^{ \infty } d \tau e^{i \omega \tau} 
\quad \
\label{a07d} \\
\int dE  f_{ }\left( E \right) 
e^{ \frac{ i }{ \hbar  } E   \tau  } 
\int dE ^{\prime}  f_{ }\left( E ^{\prime} \right) 
e^{ \frac{ -i }{ \hbar  } E ^{\prime}  \tau }
\nonumber \\
\sum_{n,m}^{} 
e ^{i \Omega  n \tau}
e ^{i \Omega t \left( n - m  \right)}
S_{F}^{*}\left(E_{n},E  \right)
S_{F}\left(E_{m},E  \right)
\nonumber \\
= -
\int dE  
\sum_{n}^{} 
f_{ }\left( E \right) 
f_{ }\left( E_{n} + \hbar \omega \right) 
\nonumber \\
S_{F}^{*}\left(E_{n},E  \right)
S_{F}\left(E_{n + \ell},E  \right) .
\nonumber 
\end{eqnarray}
\ \\ \noindent
This is the same as $B_{2}^{}$, Eq.~(\ref{a03b}). 

And finally, let us calculate the Fourier transform of $A_{1}$, Eq.~(\ref{a06b}),

\begin{eqnarray}
A_{1, \ell}\left( \omega \right) = 
\frac{ 1 }{ h }
\int\limits _{0}^{ {\cal T} _{0} } \frac{dt  }{ {\cal T} _{0} } e^{i \Omega \ell t} 
\int\limits _{- \infty}^{ \infty } d \tau e^{i \omega \tau}
\int dE  f_{ }\left( E \right) 
e^{ \frac{ i }{ \hbar  } E   \tau  } \ \ \ \ 
\label{a07e} \\
\int dE ^{\prime}  f_{ }\left( E ^{\prime} \right) 
e^{ \frac{ -i }{ \hbar  } E ^{\prime}  \tau }
\sum_{n,m,j,k}^{} 
e ^{i \Omega  (n - j) \tau}
e ^{i \Omega t \left( n + k  - m -j \right)}
\nonumber \\
S_{F}^{*}\left(E_{n},E  \right)
S_{F}\left(E_{m},E  \right)
S_{F}^{}\left(E_{j} ^{\prime},E ^{\prime}  \right)
S_{F}^{*}\left(E_{k} ^{\prime},E ^{\prime}  \right) 
\nonumber \\
= 
\int dE  
\sum_{n,m,j}^{} 
f_{ }\left( E \right) 
f_{ }\left( E_{n-j} + \hbar \omega \right) 
\nonumber \\
S_{F}^{*}\left(E_{n},E  \right)
S_{F}\left(E_{m},E  \right)
S_{F}^{}\left(E_{n} + \hbar \omega,E_{n-j} + \hbar \omega  \right)
\nonumber \\
S_{F}^{*}\left(E_{m - \ell} + \hbar \omega,E_{n-j} + \hbar \omega \right) .
\nonumber 
\end{eqnarray}
\ \\ \noindent
We denote $q = n - j$ instead of $j$, and get, 

\begin{eqnarray}
A_{1, \ell}\left( \omega \right) = 
\int dE  
\sum_{q}^{} 
f_{ }\left( E \right) 
f_{ }\left( E_{q} + \hbar \omega \right) \ \ \ \ 
\label{a07f} \\
\sum_{n,m}^{} 
S_{F}^{*}\left(E_{n},E  \right)
S_{F}\left(E_{m},E  \right)
\nonumber \\
S_{F}^{*}\left(E_{m - \ell} + \hbar \omega,E_{q} + \hbar \omega \right) 
S_{F}^{}\left(E_{n} + \hbar \omega,E_{q} + \hbar \omega  \right) .
\nonumber 
\end{eqnarray}
\ \\ \noindent
After the shift $m- \ell \to m$, we find that this is nothing but $B_{1}^{}$, Eq.~(\ref{a03a}). 
Therefore, Eq.~(\ref{a05}) indeed holds. 

The proof of Eq.~(\ref{n02}) is completed. 
\end{subequations}

\subsection{The Fourier transformation for the product of first-order correlation functions}

\label{ftg1} 

Here I show how the second term on the right hand side of Eq.~(\ref{n03}) is calculated from the corresponding term in Eq.~(\ref{n02}) in the limit of $ {\cal T} _{0} \to \infty$. 

The current generated by a periodically driven source is periodic in time, $I_{ \alpha}(t) = I_{ \alpha}(t + {\cal T} _{0})$, $ \alpha = 3,4$. 
Therefore, we can expand it into the Fourier series,

\begin{eqnarray}
I_{ \alpha}(t) &=& \sum\limits_{n=-\infty}^{\infty} e^{- in \Omega t} I_{ \alpha, n}, 
\nonumber \\
\label{a08} \\
I_{ \alpha, n} &=& \int _{0}^{ {\cal T} _{0} } \frac{ dt }{ {\cal T} _{0} } e^{i n \Omega t} .
\nonumber 
\end{eqnarray}
\ \\ \noindent
In the limit of $ {\cal T} _{0} \to \infty$, we introduce a continuous frequency $ \omega_{n} = n \Omega$, replace $ \sum_{n=-\infty}^{\infty} \to \int _{}^{ } d \omega_{n}/ \Omega$, and introduce the continuous Fourier transformation, 

\begin{eqnarray}
I_{ \alpha}\left( \omega_{n} \right) &=& \int _{- \infty}^{ \infty } dt e^{i \omega_{n} t} I_{ \alpha}\left( t \right),
\nonumber \\
\label{a09} \\
 I_{ \alpha}\left( t \right) &=& \int _{- \infty}^{ \infty } \frac{d \omega_{n} }{ \Omega } e^{-i \omega_{n} t} I_{ \alpha}\left( \omega_{n} \right) .
\nonumber 
\end{eqnarray}
\ \\ \noindent
Then we use the second line of above equation in Eq.~(\ref{n02}) and get the corresponding term in Eq.~(\ref{n03}).

\section{The Fourier transform of the function $G ^{(2)}$ for a $2$-electron leviton}
\label{ap02}

The first-order correlation function of a $2$-electron leviton,  $G_{N=2} ^{(1)}$, is given in Eqs.~(\ref{L02}) and (\ref{L03}) for $N=2$. 
The corresponding second-order correlation function reads, 

\begin{eqnarray}
G_{N=2}^{(2)}\left(t_{}+ \tau, t_{}; t, t_{} + \tau \right) = 
- 
\left |  G_{N=2} ^{(1)}\left( t+ \tau; t \right)  \right |^{2}  \nonumber \\
\label{b01} \\
+
G_{N=2} ^{(1)}\left( t;t \right)  G_{N=2} ^{(1)}\left( t+ \tau; t + \tau \right) 
. 
\nonumber 
\end{eqnarray}
\ \\ \noindent
Now let us apply the Fourier transformation defined in Eq.~(\ref{a05}).

\subsection{The first term}

The Fourier transform of the first term on the right hand side of Eq.~(\ref{b01}) determines the finite-frequency noise power, 

\begin{subequations}
\label{b02}

\begin{eqnarray}
\frac{{\cal P}_{34, \ell}(\omega) }{{\cal P}_{0}  } = -
\int\limits _{0}^{ {\cal T} _{0} } dt e^{i \Omega \ell t}   
\int\limits _{- \infty}^{ \infty } d \tau e^{i \omega \tau}  
\left | v_{ \mu}  G_{N=2} ^{(1)}\left( t+ \tau; t \right)  \right |^{2} 
\nonumber \\
= -
\frac{\Gamma _{\tau}^{2} }{  \pi^{2}  }
\int\limits _{0}^{ {\cal T} _{0} } dt  e^{i \Omega \ell t}
\int\limits _{- \infty}^{ \infty } d \tau e^{i \omega \tau}
\eta^{2}\left( \frac{ \tau }{ \tau_{ \theta} } \right)
\nonumber \\
\label{b02a} \\
\Bigg\{
2\frac{ 1 }{  \left( t + \tau \right)^{2}  +  \Gamma _{\tau}^{2} }
\frac{ 1 }{  t^{2} +  \Gamma _{\tau}^{2} } 
+
\frac{1 }{ \left( t + \tau - i  \Gamma _{\tau} \right)^{2} }
\frac{1 }{ \left( t + i  \Gamma _{\tau} \right)^{2} }
\nonumber \\
+
\frac{1 }{ \left( t + \tau + i  \Gamma _{\tau} \right)^{2} }
\frac{1 }{ \left( t - i  \Gamma _{\tau} \right)^{2} }
\Bigg\}
.
\nonumber 
\end{eqnarray}

In the case when the levitons created at different periods do not overlap, $ {\cal T} _{0} \gg  \Gamma _{\tau}$, we can safely extend the limits of integration over $t$ to infinity and introduce a continuous frequency $ \omega_{ \ell} = \ell \Omega$ instead of the set of discrete frequencies $ \ell \Omega$. 
Then, to integrate over $t$, we use the following auxiliary integrals,

\begin{eqnarray}
2 \frac{\Gamma _{\tau}^{2} }{  \pi^{2}  }
\int\limits _{- \infty}^{ \infty } dt \frac{ e^{i \omega_{ \ell} t} }{  \left( t + \tau \right)^{2}  +  \Gamma _{\tau}^{2} }
\frac{ 1 }{  t^{2} +  \Gamma _{\tau}^{2} } = 
2 \frac{\Gamma _{\tau}^{} }{  \pi^{}  }
\frac{ e^{-  \left | \omega_{ \ell} \right |  \Gamma _{\tau}} }{ \tau^{2} + 4  \Gamma _{\tau}^{2} }  
\nonumber \\
\label{b02b} \\
\times 
\left\{
\begin{array}{cc}
1 + e^{-i \omega_{ \ell} \tau} - i \frac{ 2  \Gamma _{\tau} }{ \tau } \left( 1 - e^{-i \omega_{ \ell} \tau}  \right) ,
  &   \ell > 0,  \\
  &      \\
1 + e^{-i \omega_{ \ell} \tau} + i \frac{ 2  \Gamma _{\tau} }{ \tau } \left( 1 - e^{-i \omega_{ \ell} \tau}  \right) ,
  &   \ell < 0,  
\end{array}
\right.
\nonumber 
\end{eqnarray}
\ \\ \noindent
and 

\begin{eqnarray}
\frac{\Gamma _{\tau}^{2} }{  \pi^{2}  }
\int\limits _{- \infty}^{ \infty } dt 
\frac{ e^{i \omega_{ \ell} t} }{ \left( t + \tau - i  \Gamma _{\tau} \right)^{2} }
\frac{1 }{ \left( t + i  \Gamma _{\tau} \right)^{2} }
= 
\frac{\Gamma _{\tau}^{2} }{  \pi^{2}  }
\frac{ e^{-  \left | \omega_{ \ell} \right |   \Gamma _{\tau}} }{ \left( \tau^{2} + 4  \Gamma _{\tau}^{2} \right)^{2} }  
\nonumber \\
\label{b02c} \\
\left\{
\begin{array}{cc}
e^{-i \omega_{ \ell} \tau}
\bigg[
- 2 \pi \omega_{ \ell} \left(  \tau^{2} - 4  \Gamma _{\tau}^{2} + 4 i  \Gamma _{\tau} \tau \right) \\
+ \frac{ 4 \pi \left( 8  \Gamma _{\tau}^{3} - 6  \Gamma _{\tau} \tau^{2} \right) + 4 \pi i \left(  \tau^{3} - 12  \Gamma _{\tau}^{2} \tau \right) }{ \tau^{2} + 4  \Gamma _{\tau}^{2} }
\bigg]
,
  &   \ell > 0,  \\
  &      \\
 2 \pi \omega_{ \ell} \left(  \tau^{2} - 4  \Gamma _{\tau}^{2} + 4 i  \Gamma _{\tau} \tau \right) \\ \\
+ \frac{ 4 \pi \left( 8  \Gamma _{\tau}^{3} - 6  \Gamma _{\tau} \tau^{2} \right) + 4 \pi i \left(  \tau^{3} - 12  \Gamma _{\tau}^{2} \tau \right) }{ \tau^{2} + 4  \Gamma _{\tau}^{2} },
  &   \ell < 0. 
\end{array}
\right.
\nonumber 
\end{eqnarray}
\ \\ \noindent
The noise power becomes,

\begin{eqnarray}
\frac{{\cal P}_{34, \ell}(\omega) }{{\cal P}_{0}  } = -
e^{-  \left | \omega_{ \ell} \right |   \Gamma _{\tau}}
\frac{   2 \Gamma _{\tau} }{ \pi }
\int\limits _{- \infty}^{ \infty } d \tau 
\frac{ \eta^{2}\left( \frac{ \tau }{ \tau_{ \theta} } \right) }{ \tau^{2} + 4  \Gamma _{\tau}^{2} } 
\Bigg\{
\quad\quad\quad
\label{b02d} \\ 
e^{i \omega \tau} + 
e^{i \tau \left( \omega - \omega_{ \ell} \right)} 
- i\frac{ 2  \Gamma _{\tau} }{ \tau }  {\rm sgn}( \omega_{ \ell})  \left[ 
e^{i \omega \tau} - 
e^{i \tau \left( \omega - \omega_{ \ell} \right)}  
\right]  
\nonumber \\
+ 
\frac{ e^{i \tau \left( \omega - \omega_{ \ell} \right)} A + e^{i \omega \tau}A^{*}  }{ \tau^{2} + 4  \Gamma _{\tau}^{2} } 
\Bigg\} ,
\nonumber \\
\nonumber \\
A =    \Gamma _{\tau}  \left | \omega_{ \ell} \right | \left ( 4  \Gamma _{\tau}^{2} - \tau^{2} \right) - i 4   \Gamma _{\tau}^{2} \omega_{ \ell}  \tau 
\nonumber \\
\nonumber \\
+ 2 \Gamma _{\tau}  \frac{  \left( 8  \Gamma _{\tau}^{3} - 6  \Gamma _{\tau} \tau^{2} \right) + i\,  {\rm sgn}( \omega_{ \ell}) \left(  \tau^{3} - 12  \Gamma _{\tau}^{2} \tau \right) }{ \tau^{2} + 4  \Gamma _{\tau}^{2} } .
\nonumber  
\end{eqnarray}
\ \\ \noindent
This equation satisfies the general symmetry properties, $ {\cal P}_{ \ell}\left(  \omega \right) = {\cal P}_{ -\ell}^{*}\left(- \omega \right)$ and $ {\cal P}_{ \ell}\left(  \omega \right) = {\cal P}_{ \ell}\left( \ell \Omega - \omega \right)$. \cite{Moskalets:2009dk}
To prove the latter one we need to change $ \tau \to - \tau$.  
Moreover, easy to see, that Eq.~(\ref{b02d}) is real, 

\begin{eqnarray}
\frac{{\cal P}_{34, \ell}(\omega) }{{\cal P}_{0}  } &=& - 
e^{-  \left | \omega_{ \ell} \right |   \Gamma _{\tau}}
\frac{ 4  \Gamma _{\tau} }{ \pi }
\int\limits _{0}^{ \infty } d \tau 
\frac{ \eta^{2}\left( \frac{ \tau }{ \tau_{ \theta} } \right) }{ \tau^{2} + 4  \Gamma _{\tau}^{2} } 
\nonumber \\
&&
\times
\Bigg\{
B \left [ \cos \left( \omega \tau \right) + \cos \left( \left [ \omega - \omega_{ \ell}\right]  \tau \right) \right]
\nonumber \\
&&
+
C\left [ \sin\left( \omega \tau \right) - \sin\left( \left [ \omega - \omega_{ \ell} \right] \tau \right)
 \right]
\Bigg\} ,
\nonumber \\
\label{b02e} \\ 
B &=& 
1 
+  
\Gamma _{\tau} \left | \omega_{ \ell} \right | \frac{  4  \Gamma _{\tau}^{2}  - \tau^{2}  }{ \tau^{2} + 4  \Gamma _{\tau}^{2} } 
+ 
4  \Gamma _{\tau}^{2} \frac{ 4  \Gamma _{\tau}^{2} - 3  \tau^{2} }{ \left( \tau^{2} + 4  \Gamma _{\tau}^{2} \right)^{2} }
,
\nonumber \\
C &=&  \frac{ 2  \Gamma _{\tau} }{ \tau }  {\rm sgn}( \omega_{ \ell}) 
- 4  \Gamma _{\tau}^{2}  \frac{   \omega_{ \ell} \tau  }{ \tau^{2} + 4  \Gamma _{\tau}^{2} } 
\nonumber \\
&&
+ 
2  \Gamma _{\tau} {\rm sgn}( \omega_{ \ell}) \frac{ \tau^{3} - 12  \Gamma _{\tau}^{2} \tau  }{ \left( \tau^{2} + 4  \Gamma _{\tau}^{2} \right)^{2} }
.
\nonumber  
\end{eqnarray}
\ \\ \noindent
Note, the above equation is the total cross-correlation noise power, not the excess noise power, which is more convenient in the case of the auto-correlation noise measurement, see, e.g., Ref.~\onlinecite{Parmentier:2012ed}. 

The noise power ${\cal P}_{34, \ell}$, Eq.~(\ref{b02e}), vanishes at large frequencies, $ \omega, \omega_{ \ell} \gg \Gamma _{\tau}^{-1}$. 
This fact tells us that quantum noise\cite{Gardiner2000}, which grows with frequency, is not manifested here.  

\end{subequations}

\subsection{The second term}

I denote the Fourier transform of the second line of Eq.~(\ref{b01}) as  $G ^{(2),cl}$.
Since this part survives at high temperatures, which is expected for the classical contribution, I introduce the superscript ``$cl$''. 
This part is expressed in terms of the Fourier transform of a current, carried by the levitons, $I(t) = e v_{ \mu}  G_{N=2} ^{(1)}\left( t;t \right)$, as follows, 

\begin{subequations}
\label{b03}

\begin{eqnarray}
G ^{(2),cl}_{ \ell}\left( \omega \right) &=& \frac{ 1 }{ e^{2} v_{ \mu}^{2} }
\int\limits _{0}^{ {\cal T} _{0} } e^{i \Omega \ell t} dt  
\int\limits _{- \infty}^{ \infty } d \tau e^{i \omega \tau}
I(t) I(t+ \tau).
\nonumber \\
\label{b03a}
\end{eqnarray}
\ \\ \noindent
Then I use the periodicity condition and write, $I(t) = \sum_{n=-\infty}^{\infty} e^{ - i n \Omega t} I_{n}$. 
The above equation becomes, 

\begin{eqnarray}
G ^{(2),cl}_{ \ell}\left( \omega \right) &=& \frac{ 2 \pi  {\cal T} _{0} }{ e^{2} v_{ \mu}^{2} } \sum\limits_{n=-\infty}^{\infty}  \delta\left(  \omega - n\Omega \right) I_{n} I_{ \ell - n}. 
\nonumber \\ 
\label{b03b}
\end{eqnarray}
\ \\ \noindent
In the long period limit, when the levitons emitted at different periods do not overlap, I introduce a continuous frequency $ \omega_{n} = n \Omega$ and replace the sum by the integral,

\begin{eqnarray}
\sum\limits_{n=-\infty}^{\infty} \to \int\limits _{ - \infty}^{ \infty } \frac{ d \omega_{n} }{ \Omega } .
\label{b03c}
\end{eqnarray}
\ \\ \noindent
Correspondingly, the coefficients of a discrete Fourier transformation are replaces by the coefficients of a continuous Fourier transformation, $I_{n} \to I\left( \omega_{n} \right)/ {\cal T} _{0}$. 
Then the equation (\ref{b03b}) becomes,

\begin{eqnarray}
v_{ \mu}^{2} G ^{(2),cl}_{ \ell}\left( \omega \right) &=& \frac{ {\cal T} _{0}^{2} }{ e^{2}  } I\left( \omega \right) I\left( \omega_{ \ell} - \omega \right) ,
\label{b03d}
\end{eqnarray}
\ \\ \noindent
with $I\left( \omega \right) = e v_{ \mu} G_{N=2}^{(1)}\left(  \omega \right)$ and $G_{N=2}^{(1)}\left(  \omega \right) = 2 e^{- \left | \omega \right |  \Gamma _{\tau}}$. 
\end{subequations}

According to Eq.~(\ref{n03}), the sum of Eqs.~(\ref{b02e}) and (\ref{b03d}) defines the Fourier transform of the second-order correlation function,

\begin{eqnarray}
v_{ \mu}^{2} G_{N=2, \ell} ^{(2)}\left(  \omega \right) = \frac{{\cal P}_{34, \ell}(\omega) }{{\cal P}_{0}  } + v_{ \mu}^{2} G ^{(2),cl}_{ \ell}\left( \omega \right). 
\label{b04}
\end{eqnarray}
\indent 
The above function is what is plotted in Fig.~\ref{fig3}.

\end{document}